\documentclass[journal]{IEEEtran}

\usepackage{cite}
\usepackage[pdftex]{graphicx}
\usepackage{array}
\usepackage{url}
\usepackage{multirow}
\usepackage{multicol}
\usepackage{booktabs}
\usepackage{makecell}
\usepackage{threeparttable}
\usepackage{xcolor}
\usepackage{algorithm}
\usepackage{algorithmic}
\usepackage{amsmath}
\usepackage{amsmath, bm}
\usepackage{amsfonts}
\usepackage{amsthm}
\usepackage{bm}
\usepackage{mathrsfs}
\usepackage{enumerate}
\usepackage{nomencl}
\makenomenclature
\usepackage{etoolbox}

\begin{document}

\title{Network-Constrained Unit Commitment with Flexible Temporal Resolution}

\author{Zekuan~Yu,~\IEEEmembership{Graduate~Student~Member,~IEEE,}
    Haiwang~Zhong,~\IEEEmembership{Senior~Member,~IEEE,}
    Guangchun~Ruan,~\IEEEmembership{Member,~IEEE,}
    and Xinfei~Yan,~\IEEEmembership{Student~Member,~IEEE} 
\thanks{The work of Z. Yu, H. Zhong and X. Yan was supported in part by the National Key R\&D Program of China under Grant 2022YFB2403500, and in part by the National Natural Science Foundation of China under Grant 52122706. \textit{(Corresponding author: Haiwang Zhong.)}}
\thanks{Z. Yu, H. Zhong and X. Yan are with the State Key Laboratory of Power System Operation and Control, Dept. of Electrical Engineering, Tsinghua University, Beijing, China (e-mail: zhonghw@tsinghua.edu.cn). H. Zhong is also with the Sichuan Energy Internet Research Institute, Tsinghua University.}
\thanks{G. Ruan is with the Lab for Information \& Decision Systems, Massachusetts Institute of Technology, Massachusetts, U.S. (email: gruan@mit.edu).}}

\maketitle

\begin{abstract}
Modern network-constrained unit commitment (NCUC) bears a heavy computational burden due to the ever-growing model scale. This situation becomes more challenging when detailed operational characteristics, complicated constraints, and multiple objectives are considered. We propose a novel simplification method to determine the flexible temporal resolution for acceleration and near-optimal solutions. The flexible temporal resolution is determined by analyzing the impact on generators in each adaptive time period with awareness of congestion effects. Additionally, multiple improvements are employed on the existing NCUC model compatible with flexible temporal resolution to reduce the number of integer variables while preserving the original features. A case study using the IEEE 118-bus and the Polish 2736-bus systems verifies that the proposed method achieves substantial acceleration with low cost variation and high accuracy.
\end{abstract}

\begin{IEEEkeywords}
flexible temporal resolution, acceleration, congestion, power system operation, energy management, day-ahead scheduling
\end{IEEEkeywords}



\renewcommand\nomgroup[1]{%
    \item[\bfseries
    \ifstrequal{#1}{C}{Common parameters}{%
    \ifstrequal{#1}{T}{Flexible temporal resolution determination}{%
    \ifstrequal{#1}{M}{NCUC model}{}}}%
]}

\nomenclature[C]{$\tau_t$}{Starting point of adaptive time period $t$.}
\nomenclature[C]{$d_t$}{Duration of adaptive time period $t$.}
\nomenclature[C]{$T$}{Number of adaptive time periods.}
\nomenclature[C]{$T_0$}{Number of original time periods.}
\nomenclature[C]{$N_B$}{Number of buses.}
\nomenclature[C]{$T_{l,j}$}{Power transfer distribution factor (PTDF) of line $l$ regarding bus $j$.}
\nomenclature[C]{$b(i)$}{Bus ID of unit $i$.}
\nomenclature[C]{$\overline{P}_i$, $\underline{P}_i$}{Maximum and minimum power output of unit $i$.}
\nomenclature[C]{$N_G$}{Number of thermal units.}
\nomenclature[C]{$\Delta \tau$}{Length of an original time period.}
\nomenclature[C]{$N_L$}{Number of transmission lines.}
\nomenclature[C]{$F_l$}{Limit of power flow on line $l$.}

\nomenclature[T]{$\lambda(\cdot)$}{Impact on the generation side by the demand side.}
\nomenclature[T]{$\Delta p_i$}{Power output variation of unit $i$.}
\nomenclature[T]{$\Delta P_i$}{Boundary of power output variation of unit $i$.}
\nomenclature[T]{$\mathbf{G}_{\text{on}}$}{Set of committed units.}
\nomenclature[T]{$\Delta D_j$, $\Delta D$}{Demand variation of bus $j$ and the system.}
\nomenclature[T]{$D^\tau$}{System demand in original time period $\tau$.}
\nomenclature[T]{$\Delta pf_l$}{Power flow variation of line $l$.}
\nomenclature[T]{$\Delta PF_l$}{Power flow variation of line $l$ caused by the demand.}
\nomenclature[T]{$\mathbf{G}_+$, $\mathbf{G}_-$}{Sets of units with positive and negative PTDF values.}
\nomenclature[T]{$T_{l, \mathbf{G}_+}$, $T_{l, \mathbf{G}_-}$}{Average PTDFs of units with positive and negative PTDF values.}
\nomenclature[T]{$\Delta P_{\mathbf{G}_+}$, $\Delta P_{\mathbf{G}_-}$}{Boundaries of power output variation of units with positive and negative PTDF values.}
\nomenclature[T]{$\Delta p_{\mathbf{G}_+}$, $\Delta p_{\mathbf{G}_-}$}{Power output variations of units with positive and negative PTDF values.}
\nomenclature[T]{$\Delta \widehat{P}_{\mathbf{G}_+}$, $\Delta \widehat{P}_{\mathbf{G}_-}$}{Maximum power output variation of units with positive and negative PTDF values.}
\nomenclature[T]{$\mathbf{L}^{\tau_S, \tau_F}$}{Set of possible lines with congestion in adaptive time period $[\tau_S, \tau_F]$.}
\nomenclature[T]{$\eta_{\mathbf{G}_+}$, $\eta_{\mathbf{G}_-}$}{Proportions of power output range of units with positive and negative PTDF values.}

\nomenclature[M]{$p_i^\tau$, $p_i^{(t)}$}{Power output of unit $i$ in original time period $\tau$ and adaptive time period $t$.}
\nomenclature[M]{$u_i^\tau$, $u_i^{(t)}$}{Commitment status of unit $i$ in original time period $\tau$ and adaptive time period $t$.}
\nomenclature[M]{$C_i^{\text{OP}}$, $C_i^{\text{SU}}$}{Operational cost and startup cost of unit $i$.}
\nomenclature[M]{$\widehat{D}_j^{(t)}$, $\widehat{D}^{(t)}$}{Average demand of bus $j$ and the system in adaptive time period $t$.}
\nomenclature[M]{$\overline{D}^{(t)}$, $\underline{D}^{(t)}$}{Maximum and minimum system demand in adaptive time period $t$.}
\nomenclature[M]{$RU_D^{(t)}$, $RD_D^{(t)}$}{Maximum amount of increase and decrease of system demand in adaptive time period $t$.}
\nomenclature[M]{$r^+$, $r^-$}{Positive and negative system reserve ratio.}
\nomenclature[M]{$RU_i$, $RD_i$}{Ramping up and down rate of unit $i$.}
\nomenclature[M]{$RU_i^{(t)}$, $RD_i^{(t)}$}{Maximum amount of power output increase and decrease of unit $i$ in adaptive time periods $t-1$ and $t$.}
\nomenclature[M]{$SU_i^{(t)}$, $SD_i^{(t)}$}{Maximum power output of unit $i$ in adaptive time period $t$ after startup and before shutdown.}
\nomenclature[M]{$T(\cdot)$}{Effective range of minimum up/downtime constraints.}
\nomenclature[M]{$TU_i$, $TD_i$}{Minimum up/downtime of unit $i$.}
\nomenclature[M]{$TU_i^0$, $TD_i^0$}{Initial up/downtime of unit $i$.}
\nomenclature[M]{$\tau^\text{SU}_{t,i}$}{Turning point of power output increase in ramping up constraints of unit $i$ in adaptive time period $t$.}

\printnomenclature

\section{Introduction}
\IEEEPARstart{N}{etwork}-constrained unit commitment (NCUC) is a milestone model for power system operation, in which the controllable generators are scheduled using mixed-integer linear programming (MILP)~\cite{chen_computing_2022}. Modern power systems are becoming increasingly complicated with a number of renewable energy resources integrated into the system. In this context, the NCUC model is encountering a significant computational burden due to model scale expansion. NCUC models are mathematically formulated as MILP models whose computational complexity grows exponentially with respect to the model scale. It should be noted that the NCUC model may fail to derive a solution in a permitted time duration if no accelerations or simplifications are utilized.

A few acceleration methods have been applied to increase the computational efficiency of NCUC. For instance, heuristic algorithms are widely applied to guide the branch-and-bound (B\&B) process of NCUC~\cite{zhu_parallel_2022}. At the very beginning, warm-start algorithms \cite{parrilla_improving_2006, chen_distributed_2020} can be utilized to produce an initial feasible and integer solution to accelerate the search process by pruning branches of the B\&B tree. Based on the characteristics of thermal units and transmission or security constraints, domain knowledge can help eliminate a large number of redundant and nonbinding constraints so that the complexity of the NCUC model can be effectively reduced \cite{zhai_fast_2010, ma_redundant_2021}. A novel and simple iterative contingency-screening procedure is used in \cite{xavier_transmission_2019} to filter out 99.4\% of the redundant constraints as well. When a relaxed solution is obtained, neighborhood-search algorithms \cite{chen_high_2021} can temporarily limit the search space within its neighborhood to quickly produce a solution of high quality, and relaxation inducement algorithms \cite{bai_inducing-objective-function-based_2014,gao_internally_2022} can utilize the tendency of the relaxed values of integer variables by inducing them toward an integer solution. Another line of efforts have been made to apply machine learning or quantum-based approaches in NCUC~\cite{ruan_review_2021}. For instance, multiple machine-learning methods such as k-nearest neighbours and support vector machines are utilized to identify the nonbinding transmission constraints and to produce warm-start initial solutions \cite{xavier_learning_2021}. In \cite{morstyn_annealing-based_2023}, the author explores the applicability of the emerging quantum computing in the calculation of combinatorial optimal power flow problems such as NCUC. Deep neural networks and quantum computing are embedded with an advanced Surrogate Lagrangian Relaxation (SLR) framework in \cite{wu_synergistic_2023} and \cite{feng_novel_2023} to accelerate the computation of subproblems.
All these algorithms aim to improve the efficiency of solving NCUC from the perspective of the designs of the calculation process, but their performance may depend on specific settings and may sometimes appear mediocre.

The simplification of NCUC models can also accelerate computing by increasing the model compactness~\cite{chen_security-constrained_2022}, and the idea is to reduce the number of integer variables directly in the model setup. Methods of this kind can generally be classified into two categories: unit dimension and temporal dimension. 

The simplification regarding the unit dimension is known as unit clustering, where thermal units with the same features and parameters are not distinguished from each other \cite{hara_method_1966}. The authors in \cite{palmintier_heterogeneous_2014} and \cite{langrene_dynamic_2011} extended this method to include similar yet not the same thermal units, where a single integer variable is assigned to represent the commitment status of each unit cluster. This method was later implemented in \cite{meus_applicability_2018}, in which the authors analyzed the errors introduced by aggregating nonidentical units and proposed an approach to obtaining the commitment statuses of all the units in each cluster while guaranteeing feasibility. In \cite{du_high-efficiency_2019}, the authors managed to introduce network constraints into NCUC with unit clustering and verified the effectiveness of this method for power system planning studies.

Flexible temporal resolution is another promising option for acceleration. The idea originates from the fact that the unit statuses are very likely to remain the same if the operating conditions are not changed in a certain range of time. In this case, several temporal intervals can be locally combined to reduce the number of decision variables (both binary and continuous ones)~\cite{hoffmann_review_2020}. Such a framework was developed in \cite{pineda_time-adaptive_2019}, where a day-ahead UC model compatible with flexible temporal resolution was formulated. The determination of flexible temporal resolution is usually achieved by aggregating the time periods according to their operating conditions. In \cite{pineda_time-adaptive_2019}, the time periods were aggregated using a hierarchical clustering algorithm based on Ward's method. Several other methods for time period aggregation were applied in existing literature, including sliding window backtracking\cite{wijekoon_efficient_2018}, minimizing the sum of absolute gradients of the demand\cite{vom_stein_development_2017}, formulating a load state transition curve\cite{feng_load_2021}, etc. These methods simply aggregate the time periods according to the statistical features of the demand without considering their effect on the thermal units. The authors in \cite{zhang_enhancing_2021} developed a strategy involving the calculation of a series of small-scale NCUC models to evaluate each time period before determining the representative scheduling points so that the characteristics of thermal units can be considered at the cost of increased computational complexity. In \cite{yu_adaptive_2022}, an aggregation method combining the features of the demand and thermal units was proposed, but the effect of congestion was neglected.

In summary, existing NCUC simplification methods with respect to the temporal dimension can usually guarantee a high computational efficiency, while the accuracy and feasibility of their results may be compromised. In order to overcome this drawback, in this paper we propose a simplification method for NCUC with flexible temporal resolution, which adjusts the temporal resolution locally considering the influence of congestion. The proposed method is compatible with other acceleration methods (e.g., unit clustering), and extra acceleration is possible in a joint implementation.

The contributions of this paper are summarized as follows:

\begin{enumerate}
    \item A simplification method for NCUC is proposed to determine the flexible temporal resolution for computational acceleration with limited accuracy loss. The congestion effect is taken into consideration by analyzing the output variation of the generators.
    \item Multiple modifications are proposed and employed to enhance the existing NCUC model compatible with flexible temporal resolution to preserve the original features as much as possible. In particular, the parameters of the ramping constraints are derived by exploring the extreme circumstances concerning the original time periods.
\end{enumerate}

The remainder of this paper is organized as follows. Section~\ref{sec:frame} provides a framework overview. Section~\ref{sec:aggre} introduces the deterministic method of flexible temporal resolution. Section~\ref{sec:uc} formulates the enhanced NCUC model compatible with flexible temporal resolution. Section~\ref{sec:case} validates the proposed method using the IEEE 118-bus system and the Polish 2736-bus system. Section~\ref{sec:conclusion} draws the final conclusions.

\section{Framework} \label{sec:frame}
\begin{figure}
    \centering
    \includegraphics[width=0.45\textwidth]{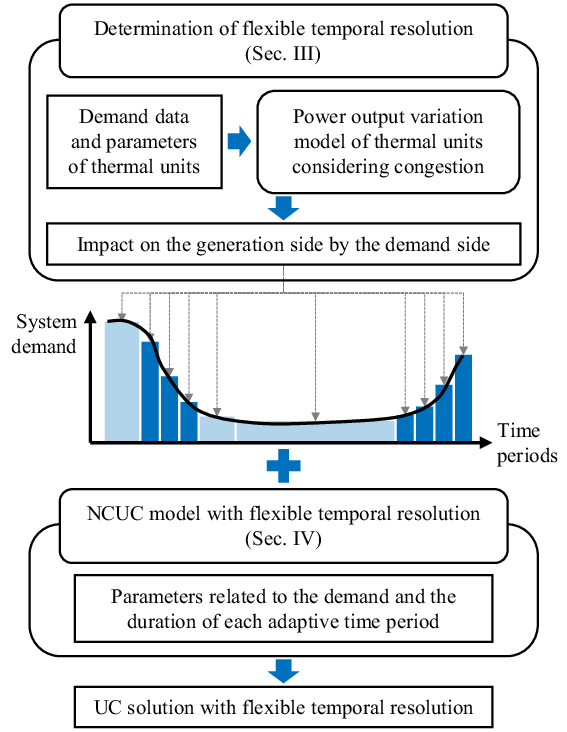}
    \caption{Framework of the proposed NCUC with flexible temporal resolution.}
    \label{fig:framework}
\end{figure}

The framework of the proposed NCUC with flexible temporal resolution is illustrated in Fig.~\ref{fig:framework}, and it is composed of two parts: determination of flexible temporal resolution and the corresponding NCUC model. The determination of flexible temporal resolution aims to estimate and to select the optimal flexible temporal resolution scheme with a given number of remaining time periods before solving the UC main problem, and the modified NCUC model is formulated to resemble the original model based on the flexible temporal resolution, both of which are dedicated to maintaining high accuracy for the proposed simplification method with high efficiency.

First, the flexible temporal resolution is determined using the demand data and the parameters of the thermal units. Here, the determination of flexible temporal resolution is transformed into the calculation of the starting point of each new time period (\textit{adaptive} time period), which adopts an optimization scheme. In this paper, we propose a generalized notion of its objective, which is to minimize the impact brought on the generation side by the demand side in each adaptive time period. To translate this notion into a formula, a power output variation model is designed to quantify and to approximate the influence of demand variation on the thermal units, which takes the effect of network congestion into consideration to enhance its accuracy. The objective is therefore defined by ensuring the feasibility of the power output variation model, and the starting points of adaptive time periods can be determined by solving the optimization model.

After the flexible temporal resolution is determined, the NCUC model is transformed to adapt to time periods with different duration. The transformed NCUC model receives the starting points of adaptive time periods and adjusts its formulation. Compared with the existing NCUC model compatible with flexible temporal resolution \cite{pineda_time-adaptive_2019}, the parameters of the constraints are modified using the characteristics of the demand regarding multiple original time periods within an adaptive time period. The parameters of the ramping constraints are specifically analyzed by deriving the corresponding extreme circumstances concerning the original time periods using the conventional ramping constraints. Moreover, two novel constraints concerning the ramping process inside an adaptive time period are added.

By combining the flexible temporal resolution and the NCUC model, a UC solution with flexible temporal resolution can therefore be obtained. Afterwards, the final power output results can be calculated by extending the UC solution to the original time period sequence and solving an economic dispatch model based on the extended commitment statuses. It is worth mentioning that, the flexible temporal resolution is classified as a simplification method of NCUC, which inevitably neglects some aspects of the original NCUC model due to the reduced number of time periods. Therefore, after being extended to the original time period sequence, this UC solution is not guaranteed to be the optimal solution, and it can even be infeasible under certain circumstances.
For instance, using the average demand in an adaptive time period in the network constraints of the NCUC model and neglecting its fluctuations may bring violations concerning the original time periods.
The correction strategy proposed in \cite{yu_adaptive_2022} is adopted in the case study to guarantee the feasibility. Nevertheless, feasibility without correction is an indication of the accuracy and the effectiveness of the proposed method for a given number of adaptive time periods. In addition, the UC solution obtained by the proposed method can be utilized for warm start in applications where theoretical optimality is necessary.

\section{Determination of Flexible Temporal Resolution}  \label{sec:aggre}

\subsection{Adaptive time period optimization model}

As mentioned above, the determination of flexible temporal resolution is formulated into an integer programming model, which optimizes the starting points of adaptive time periods in the original time period sequence:
\begin{subequations}
    \begin{align}
        \mathop {\min }\limits_{{\tau_1}, \cdots ,{\tau_T}} \sum\limits_{t = 1}^T {\lambda(\tau_t, \tau_{t+1}-1)}   \label{eq:aggre-obj}
    \end{align}
    subject to:
    \begin{align}
        & {\tau_1},{\tau_2}, \cdots {\tau_T} \in \mathbb{Z}  \label{eq:aggre-cons1}\\
        & 1 = {\tau_1} < {\tau_2} <  \cdots  < {\tau_T} \le T_0    \label{eq:aggre-cons2}
    \end{align}
\end{subequations}

In the model above, ${\tau_1}, \cdots ,{\tau_T}$ denote the starting points of $T$ adaptive time periods in the original $T_0$ time periods, and adaptive time period $t$ can therefore be expressed as $[\tau_t, \tau_{t+1}-1]$, with its duration defined as $d_t=\tau_{t+1}-\tau_t$. To satisfy the prerequisite of flexible temporal resolution, i.e., the demand should remain basically unchanged in each adaptive time period, the objective \eqref{eq:aggre-obj} is designed to minimize the summation of the impact $\lambda(\cdot)$ brought on the generation side by the variation of the demand side in all adaptive time periods, so that the influence of demand fluctuation within each adaptive time period is restrained. The derivation of $\lambda(\cdot)$ is given in the following section.

The adaptive time period optimization model \eqref{eq:aggre-obj}-\eqref{eq:aggre-cons2} is formulated as an integer programming problem. Nevertheless, the scale of this problem mainly depends on the number of original time periods $T_0$ and adaptive time periods $T$. For day-ahead NCUC, the problem scale is usually small enough so that a variety of algorithms can be adopted with acceptable time consumption. In this paper, the model is solved with dynamic programming, and the flexible temporal resolution can therefore be determined.

\subsection{Power output variation model of thermal units}

The impact brought on the generation side by the demand side is evaluated by analyzing a power output variation model for thermal units, where the power output variation of the thermal units between any two time periods $\Delta p_i$ is treated as variables. For any two time periods $\tau_s$ and $\tau_f$ within an adaptive time period $[\tau_S, \tau_F]$, the boundaries of power output variation $\pm\Delta P_i$ represent the load-following abilities of the thermal units and are considered static, mainly determined by their commitment statuses. This section aims to evaluate the impact by estimating the conditions of these boundaries for the power output variation model to be feasible. As the power generation and consumption in the entire system are balanced at any time, their variation should cancel each other out:
\begin{equation}
    \sum\limits_{i \in \mathbf{G}_{\text{on}}} \Delta p_i = \sum\limits_{j=1}^{N_B} \Delta D_j = \Delta D, \quad\forall \tau_S \le \tau_s < \tau_f \le \tau_F \label{eq:theor-bal}
\end{equation}
where $\Delta D$ denotes the system demand variation and $\mathbf{G}_{\text{on}}$ represents the set of committed thermal units in the adaptive time period.

Suppose there exists congestion on transmission line $l$ in the adaptive time period, during which the power flow reaches its upper limit $F_l$ (e.g., long-term rating). Due to congestion, the power flow on line $l$ is expected to remain the same or decrease. Therefore, the power flow variation $\Delta pf_l$ should be nonpositive:
\begin{align}
    \Delta pf_l & = \sum\limits_{i \in \mathbf{G}_{\text{on}}} T_{l,b(i)} \Delta p_i  - \sum\limits_{j = 1}^{N_B} T_{l,j} \Delta D_j \nonumber \\
    & = \sum\limits_{i \in \mathbf{G}_{\text{on}}} T_{l,b(i)} \Delta p_i + \Delta PF_l \le 0, \quad\forall \tau_S \le \tau_s < \tau_f \le \tau_F    \label{eq:theor-dpf}
\end{align}
where $T_{l,j}$ represents the power transfer distribution factor (PTDF) of line $l$ regarding bus $j$, and $b(i)$ defines the bus ID of thermal unit $i$.

Equations \eqref{eq:theor-bal} and \eqref{eq:theor-dpf} present a description of the restrictions on the power output variation of thermal units, and the existence of the adaptive time period depends on their feasibility. In order to extract their essential features, the equations are then transformed and simplified as follows.

According to the signs of the corresponding PTDF values, the thermal units in the system can be classified into two categories: $\mathbf{G}_+$ and $\mathbf{G}_-$, representing the sets of thermal units with positive and negative PTDF values. The average PTDFs of the two categories $T_{l, \mathbf{G}_+}$ and $T_{l, \mathbf{G}_-}$ are proposed and defined accordingly, where the power output ranges of thermal units are utilized as weights:
\begin{equation}
    T_{l, \mathbf{G}_{+,-}} = \frac{ \sum\limits_{i \in \mathbf{G}_{+,-}} T_{l, b(i)} \left( \overline{P}_i - \underline{P}_i \right) }{ \sum\limits_{i \in \mathbf{G}_{+,-}} \left( \overline{P}_i - \underline{P}_i \right) } \label{eq:theor-ptdf}
\end{equation}

Therefore, \eqref{eq:theor-bal} and \eqref{eq:theor-dpf} can be transformed into a simplified version:
\begin{subequations}
    \begin{align}
        & -\Delta P_{\mathbf{G}_{+,-}} \le \Delta p_{\mathbf{G}_{+,-}} \le \Delta P_{\mathbf{G}_{+,-}} \label{eq:theor-nrange} \\
        & \Delta p_{\mathbf{G}_+} + \Delta p_{\mathbf{G}_-} = \Delta D   \label{eq:theor-nbal} \\
        & T_{l, \mathbf{G}_+}\Delta p_{\mathbf{G}_+} + T_{l, \mathbf{G}_-}\Delta p_{\mathbf{G}_-} + \Delta PF_l \le 0  \label{eq:theor-ndpf}
    \end{align}
\end{subequations}
where $\Delta p_{\mathbf{G}_+}$ and $\Delta p_{\mathbf{G}_-}$ denote the total power output variation of the two types of committed units, and $\Delta P_{\mathbf{G}_+}$ and $\Delta P_{\mathbf{G}_-}$ denote the corresponding boundaries. After this transformation and simplification, the number of variables is reduced from the number of thermal units online in the adaptive time period to 2, so that the complexity is reduced for the following analysis.

Equations \eqref{eq:theor-nrange}-\eqref{eq:theor-ndpf} formulate the power output variation model. To ensure the feasibility of this model, the feasibility regions of \eqref{eq:theor-nrange}, \eqref{eq:theor-nbal} and \eqref{eq:theor-ndpf} should intersect with each other for any $\tau_S \le \tau_s < \tau_f \le \tau_F$, as illustrated in Fig.~\ref{fig:append_sim}. Therefore, the following conditions need to be satisfied (see Appendix for detailed proof):
\begin{equation}
    \begin{cases}
        \Delta P_{\mathbf{G}_+} + \Delta P_{\mathbf{G}_-} \ge \mathop {\max }\limits_{\tau_S \le \tau_s < \tau_f \le \tau_F} |\Delta D| \\
        \Delta P_{\mathbf{G}_+} \ge \Delta \widehat{P}_{\mathbf{G}_+} = \mathop {\max }\limits_{\tau_S \le \tau_s < \tau_f \le \tau_F} \frac{T_{l, \mathbf{G}_-} \Delta D + \Delta PF_l}{T_{l, \mathbf{G}_+} - T_{l, \mathbf{G}_-}}    \\
        \Delta P_{\mathbf{G}_-} \ge \Delta \widehat{P}_{\mathbf{G}_-} = \mathop {\max }\limits_{\tau_S \le \tau_s < \tau_f \le \tau_F} \frac{T_{l, \mathbf{G}_+} \Delta D + \Delta PF_l}{T_{l, \mathbf{G}_+} - T_{l, \mathbf{G}_-}}
    \end{cases} \label{eq:theor-require}
\end{equation}

\begin{figure}
    \centering
    \includegraphics[width=0.4\textwidth]{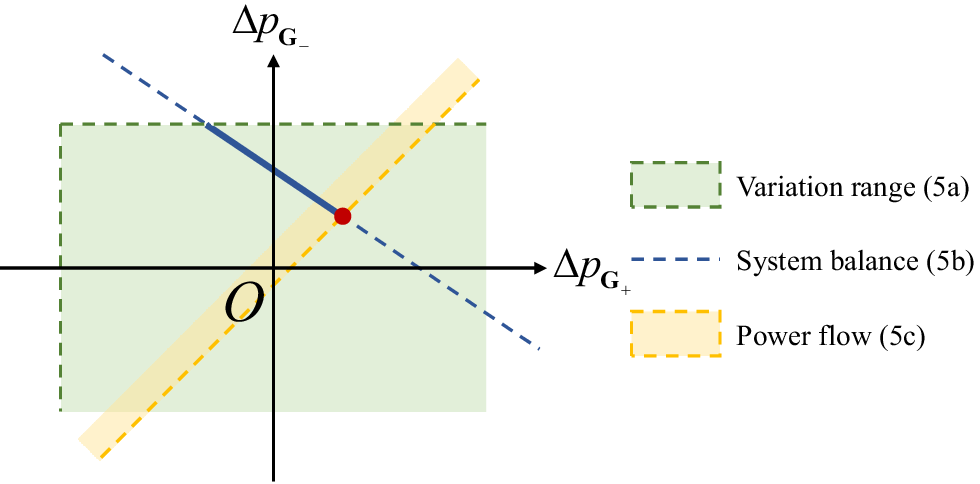}
    \caption{Illustration of the power output variation model. To ensure the feasibility of the model, the square green area \eqref{eq:theor-nrange}, the blue line \eqref{eq:theor-nbal} and the yellow half-plane \eqref{eq:theor-ndpf} must intersect.}
    \label{fig:append_sim}
\end{figure}

Equations \eqref{eq:theor-require} represent the load-following conditions for the power output variation model to be feasible. The left-hand sides of the three equations include the boundaries of the power output variation of thermal units, and the right-hand sides are composed of the variations caused by the demand side ($\Delta D$, $\Delta PF_l$). From this perspective, \eqref{eq:theor-require} can also be interpreted as the impact brought on the load-following abilities of the committed units by the demand variation. Therefore, $\lambda(\cdot)$ in \eqref{eq:aggre-obj} can be defined as follows:
\begin{equation}
    \lambda ({\tau_S},{\tau_F}) = \frac{\mathop {\max }\limits_{l \in \mathbf{L}^{\tau_S, \tau_F}} \left\{ \max |\Delta D|, \frac{\Delta \widehat{P}_{\mathbf{G}_+}}{\eta_{\mathbf{G}_+}} , \frac{\Delta \widehat{P}_{\mathbf{G}_-}}{\eta_{\mathbf{G}_-}} \right\} }{ \mathop {\max }\limits_{\tau = {\tau_S}, \cdots ,{\tau_F}} D^\tau } \label{eq:aggre-lambda}
\end{equation}
where $D^\tau$ denotes the system demand in original time period $\tau$, and $\mathbf{L}^{\tau_S, \tau_F}$ represents the possible lines with congestion in the adaptive time period, which can be estimated with various methods. For instance, they could be estimated based on historical data of system operation if it is available. In this paper, they are estimated by solving a relaxed linear programming model and analyzing the power flow, which is formulated by regarding the integer variables to be continuous and neglecting the network constraints of the original NCUC model.

The first term (range of system demand variation) in \eqref{eq:aggre-lambda} represents the general level of demand variation and is irrelevant to congestion, and the other two terms account for the influence of congestion. Due to the design and formulation of \eqref{eq:aggre-lambda}, if there is no congested line in the current adaptive time period or the congestion is mild, only the first term becomes effective, which corresponds to \cite{yu_adaptive_2022}. Otherwise, the maximum function ensures the most severely congested line to take effect in the optimization model. It is worth mentioning that the first equation in \eqref{eq:theor-require} features the load-following abilities of all the committed thermal units, including $\mathbf{G}_+$ and $\mathbf{G}_-$, while the second and third equations feature only a part of the committed units. As a result, equivalent factors $\eta_{\mathbf{G}_+}$ and $\eta_{\mathbf{G}_-}$ are introduced, which are defined as the proportion of the total power output range of the corresponding types of units:
\begin{equation}
    \eta_{\mathbf{G}_{+,-}} = \frac{\sum\limits_{i \in \mathbf{G}_{+,-}} \left( \overline{P}_i - \underline{P}_i \right)}{\sum\limits_{i = 1}^{N_G} \left( \overline{P}_i - \underline{P}_i \right)}
\end{equation}

To balance different adaptive time periods with different amounts of load-following abilities of thermal units, the maximum demand in the adaptive time period is utilized for normalization, as the total amount of online capacity is mainly determined by the maximum demand and closely related to load-following abilities. Therefore, for a certain amount of system demand variation, the corresponding time periods are more likely to be aggregated if it occurs at a higher system demand such as during the peak hours, and more likely to be separated at a lower system demand such as during the valley hours. Another feature of the adaptive time period optimization model is that, it tends to aggregate the time periods with the least demand variation and keep the rest of the original time periods rather than relatively evenly merge them. For instance, if the system demand is 100 MW, 1000 MW, 1700 MW, 2200 MW and 2500 MW in 5 consecutive time periods with no congestion, the optimal result of forming 3 adaptive time periods would be merging time periods 3-5 and keeping time periods 1 and 2 (objective: (2500-1700)/2500=0.32), instead of merging time periods 2-3 and 4-5 and keeping 1 (objective: (1700-1000)/1700+(2500-2200)/2500=0.53). Therefore, the time periods 1 and 2 with the high demand increase will not be aggregated, which corresponds to the inconsistency of commitment statuses during the ramping hours of the demand.

\section{NCUC Model with Flexible Temporal Resolution}   \label{sec:uc}

\subsection{Model formulation}

With flexible temporal resolution, the conventional NCUC model formulation needs to be modified to adapt to time periods with different duration, with a reduction in integer variables while maintaining most of the original characteristics. The modified NCUC model with flexible temporal resolution is formulated as follows, which is based on the model proposed in \cite{pineda_time-adaptive_2019}:
\begin{subequations}
    \begin{equation}
        \mathop {\min }\limits_{(u_i^{(t)},p_i^{(t)})} \sum\limits_{t = 1}^T {\sum\limits_{i = 1}^{{N_G}} {C_i^{\text{OP}}(p_i^{(t)},u_i^{(t)}){d_t}\Delta \tau + C_i^{\text{SU}}(u_i^{(t)},u_i^{(t-1)})} } \label{eq:uc-obj}
    \end{equation}
    subject to:
    \begin{align}
        & \sum\limits_{i = 1}^{{N_G}} {p_i^{(t)}}  = \widehat{D}^{(t)}, \quad \forall t \label{eq:uc-con-balance} \\
        & \sum\limits_{i = 1}^{{N_G}} {u_i^{(t)}\overline{P}_i }  \ge (1 + {r^ + })\overline{D}^{(t)} , \quad \forall t \label{eq:uc-con-resup} \\
        & \sum\limits_{i = 1}^{{N_G}} {u_i^{(t)}\underline{P}_i }  \le (1 - {r^ - })\underline{D}^{(t)} , \quad \forall t \label{eq:uc-con-resdn} \\
        & \sum\limits_{i = 1}^{{N_G}} {u_i^{(t)}R{U_i}\Delta \tau}  \ge RU_D^{(t)}, \quad \forall t \label{eq:uc-con-resrup} \\
        & \sum\limits_{i = 1}^{{N_G}} {u_i^{(t)}R{D_i}\Delta \tau}  \ge RD_D^{(t)}, \quad \forall t \label{eq:uc-con-resrdn} \\
        &  - {F_l} \le \sum\limits_{i = 1}^{{N_G}} {T_{l,b(i)} p_i^{(t)}}  - \sum\limits_{j = 1}^{{N_B}} {T_{l,j} \widehat{D}_j^{(t)}}  \le {F_l}, \nonumber \\
        & \quad \quad \quad \forall l = 1, \cdots ,{N_L},\forall t \label{eq:uc-con-net} \\
        & u_i^{(t)}\underline{P}_i  \le p_i^{(t)} \le u_i^{(t)}\overline{P}_i , \quad \forall i,\forall t \label{eq:uc-con-range} \\
        & p_i^{(t)} - p_i^{(t-1)} \le RU_i^{(t)}u_i^{(t-1)} + SU_i^{(t)}(u_i^{(t)} - u_i^{(t-1)}) \nonumber \\
        & \quad \quad \quad + \overline{P}_i (1 - u_i^{(t)}), \quad \forall i,\forall t \label{eq:uc-con-rampup} \\
        & p_i^{(t-1)} - p_i^{(t)} \le RD_i^{(t)}u_i^{(t)} - SD_i^{(t-1)}(u_i^{(t)} - u_i^{(t-1)}) \nonumber \\
        & \quad \quad \quad + \overline{P}_i (1 - u_i^{(t-1)}), \quad \forall i,\forall t \label{eq:uc-con-rampdn} \\
        & u_i^{(k)} \ge u_i^{(t)} - u_i^{(t-1)}, \quad \forall i,\forall t \le k \le T(t,T{U_i}),\forall t \label{eq:uc-con-minup} \\
        & u_i^{(k)} \le 1 - (u_i^{(t-1)} - u_i^{(t)}), \quad \forall i,\forall t \le k \le T(t,T{D_i}),\forall t \label{eq:uc-con-mindn} \\
        & u_i^{(k)} = 1,\quad \forall 1 \le k \le T(1,T{U_i} - TU_i^0), \nonumber \\
        & \quad \quad \quad \text{if}\;u_i^{(0)} = 1,TU_i^0 < T{U_i} \label{eq:uc-con-minup0} \\
        & u_i^{(k)} = 0,\quad \forall 1 \le k \le T(1,T{D_i} - TD_i^0), \nonumber \\
        & \quad \quad \quad \text{if}\;u_i^{(0)} = 0,TD_i^0 < T{D_i} \label{eq:uc-con-mindn0}
    \end{align}
\end{subequations}

Unless otherwise noted, all $\forall t$ in the equations above indicate $\forall t = 1, \cdots, T$, and all $\forall i$ indicate $\forall i = 1, \cdots, N_G$.

In the model above, $p_i^{(t)}$ and $u_i^{(t)}$ represent the power output and commitment status of thermal unit $i$ in adaptive time period $t$. The objective \eqref{eq:uc-obj} is to minimize the total operational cost $C_i^{\text{OP}}$ and the startup cost $C_i^{\text{SU}}$ over the time horizon, which is adjustable according to different scheduling goals. Similar to conventional NCUC models, the constraints of the proposed model include system balance constraints \eqref{eq:uc-con-balance}, reserve constraints \eqref{eq:uc-con-resup}-\eqref{eq:uc-con-resdn}, network constraints \eqref{eq:uc-con-net}, power output ranges \eqref{eq:uc-con-range}, ramping constraints \eqref{eq:uc-con-rampup}-\eqref{eq:uc-con-rampdn} and minimum up/downtime constraints \eqref{eq:uc-con-minup}-\eqref{eq:uc-con-mindn0}. For the minimum up/downtime constraints, their effective range $[t,T(t,DT)]$ is determined by the duration of the time periods, reflected with a function $T(\cdot)$. For instance, assume the unit switches to online in adaptive time period $t$ and is required to stay online in at least 4 consecutive original time periods. If the adaptive time period $t$, $t+1$ and $t+2$ contains 2, 3 and 4 original time periods respectively, the effective range of the minimum uptime constraint here should be $[t,t+1]$. As explained above, $T(\cdot)$ represents the first time period after $t$ which satisfies that the length of $[t,T(t,DT)]$ is no smaller than $DT$:
\begin{align}
    T(t,DT) = \begin{cases}
        \mathop {\min }\limits_{t' \ge t} t' & \text{s.t.} \sum\limits_{t'' = t}^{t'} {{d_{t''}}\Delta \tau}  \ge DT, \\
        & \sum\limits_{t'' = t}^T {{d_{t''}}\Delta \tau}  \ge DT \\
        T, & \sum\limits_{t'' = t}^T {{d_{t''}}\Delta \tau}  < DT
    \end{cases}
\end{align}

As more system-wide constraints are included in the proposed model than \cite{pineda_time-adaptive_2019}, their parameter values are adjusted to be distinguished from conventional NCUC models, mainly reflected in the variety of forms of demand data. For system balance constraints and network constraints, the average demand in the time period is utilized, while its maximum and minimum are used in reserve constraints. In addition, a new type of constraints, i.e., ramping reserve constraints \eqref{eq:uc-con-resrup}-\eqref{eq:uc-con-resrdn}, is added to ensure the ramping capabilities of committed units facing the fluctuation of demand within the time period. For these constraints, the maximum amount of increase and decrease of system demand in an adaptive time period are adopted.

\subsection{Transformation of ramping constraints}

\begin{figure}
    \centering
    \includegraphics[width=0.48\textwidth]{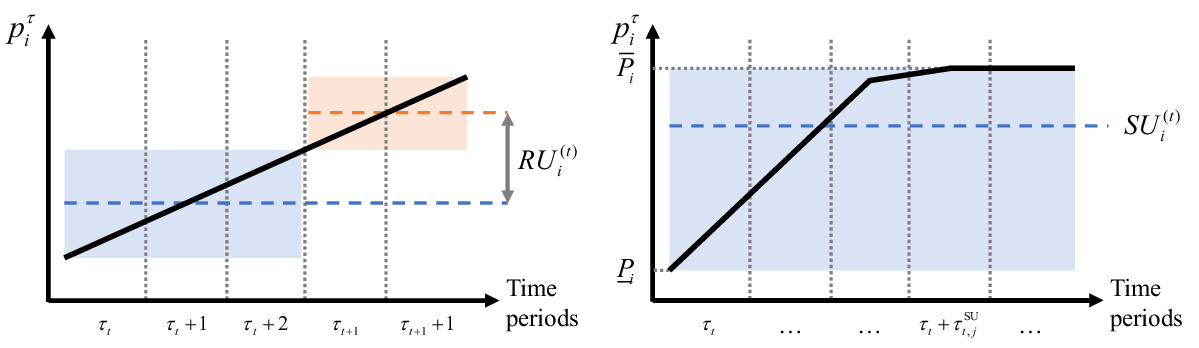}
    \caption{Illustration of the extreme circumstances concerning the original time periods for the ramping constraints. For $RU_i^{(t)}$, the power output of the unit increases with the maximum ramping capability. For $SU_i^{(t)}$, the power output of the unit increases with the maximum ramping capability from $\underline{P}_i$ to $\overline{P}_i$.}
    \label{fig:ramping}
\end{figure}

Apart from those mentioned above, the major difference between the NCUC model with flexible temporal resolution and the conventional NCUC models lies in the ramping constraints \eqref{eq:uc-con-rampup}-\eqref{eq:uc-con-rampdn}, as they represent the temporal coupling characteristics of thermal units. Taking \eqref{eq:uc-con-rampup} as an example, $RU_i^{(t)}$ denotes the maximum amount of power output increase if unit $i$ remains online in time periods $t-1$ and $t$, and $SU_i^{(t)}$ denotes the maximum power output if unit $i$ switches from offline in $t-1$ to online in $t$. These need to be determined according to the extreme circumstances concerning the original time periods, as illustrated in Fig.~\ref{fig:ramping}. In particular, the value of $SU_i^{(t)}$ (and $SD_i^{(t)}$) in this paper incorporates a detailed analysis on the extreme circumstances, which differs from \cite{pineda_time-adaptive_2019}.

First, the conventional ramping up constraint is given as follows:
\begin{equation}
    p_i^\tau - p_i^{\tau-1} \le RU_i \Delta \tau u_i^{\tau-1} + \underline{P}_i(u_i^\tau - u_i^{\tau-1}) + \overline{P}_i (1 - u_i^\tau) \label{eq:conv-con-rampup}
\end{equation}
where the first parameter $RU_i \Delta \tau$ represents the maximum amount of power output increase between two original time periods. In order to derive $RU_i^{(t)}$, suppose the thermal unit is online in adaptive time periods $t-1$ and $t$, i.e., from $\tau_{t-1}$ to $\tau_{t+1}-1$. Due to \eqref{eq:conv-con-rampup}, the following equations hold:
\begin{align}
    \begin{cases}
        p_i^\tau \ge p_i^{\tau_t-1} - (\tau_t-1 - \tau) RU_i \Delta \tau, & \forall \tau_{t-1} \le \tau \le \tau_t-1  \\
        p_i^\tau \le p_i^{\tau_t} + (\tau - \tau_t) RU_i \Delta \tau, & \forall \tau_t \le \tau \le \tau_{t+1}-1   \\
        p_i^{\tau_t} - p_i^{\tau_t-1} \le RU_i \Delta \tau
    \end{cases}
\end{align}

Using the average amount of power output in the adaptive time period as a representative, the upper bound of the difference between $p_i^{(t)}$ and $p_i^{(t-1)}$ can be calculated, and $RU_i^{(t)}$ can therefore be defined:
\begin{align}
    p_i^{(t)} - p_i^{(t-1)} & \le \frac{ \sum\limits_{\tau = \tau_t}^{\tau_{t+1}-1} p_i^{\tau_t} + (\tau - \tau_t) RU_i \Delta \tau }{ d_t } \nonumber   \\
    & - \frac{ \sum\limits_{\tau = \tau_{t-1}}^{\tau_t-1} p_i^{\tau_t-1} - (\tau_t-1 - \tau) RU_i \Delta \tau }{d_{t-1}}   \nonumber \\
    & = p_i^{\tau_t} - p_i^{\tau_t-1} + \left( \frac{d_t + d_{t-1}}{2} - 1 \right) RU_i \Delta \tau   \nonumber \\
    & \le \frac{d_t + d_{t-1}}{2} RU_i \Delta \tau = RU_i^{(t)} \label{eq:trans-rampup}
\end{align}

Equation \eqref{eq:trans-rampup} suggests that, if the unit is online in $t-1$ and $t$, $p_i^{(t)} - p_i^{(t-1)}$ reaches its maximum when the power output increases at its ramping up rate $RU_i$ in all the original time periods of $t-1$ and $t$.

In \eqref{eq:conv-con-rampup}, the second parameter $\underline{P}_i$ represents the maximum power output after switching from offline to online, which is assumed to be its minimum in this paper. Suppose the same startup process occurs in adaptive time periods $t-1$ and $t$, i.e., the unit remains offline in $[\tau_{t-1}, \tau_t-1]$ and becomes online in $[\tau_t, \tau_{t+1}-1]$. Due to \eqref{eq:conv-con-rampup}, the following equation holds:
\begin{equation}
    p_i^{\tau} \le \min \{ \underline{P}_i + (\tau - \tau_t) RU_i \Delta \tau , \overline{P}_i\}, \quad \forall \tau_t \le \tau \le \tau_{t+1} - 1
\end{equation}

The right-hand side of the equation above is a piecewise linear function of $\tau$ with two segments. By specifying the turning point $\tau^\text{SU}_{t,i}$, it can be transformed as follows:
\begin{equation}
    \begin{cases}
        p_i^{\tau} \le \underline{P}_i + (\tau - \tau_t) RU_i \Delta \tau, & \forall \tau_t \le \tau \le \tau_t + \tau^\text{SU}_{t,i} - 1 \\
        p_i^{\tau} \le \overline{P}_i, & \forall \tau_t + \tau^\text{SU}_{t,i} \le \tau \le \tau_{t+1} - 1
    \end{cases}
\end{equation}

Similarly, using the average amount of power output in the adaptive time period as a representative, the upper bound of $p_i^{(t)}$ can be calculated, and $SU_i^{(t)}$ can therefore be defined as follows:
\begin{align}
    p_i^{(t)} & \le \frac{ \sum\limits_{\tau = \tau_t}^{\tau_t + \tau^\text{SU}_{t,i} - 1} \underline{P}_i + (\tau - \tau_t) RU_i \Delta \tau }{ d_t } + \frac{ \sum\limits_{\tau = \tau_t + \tau^\text{SU}_{t,i}}^{\tau_{t+1} - 1} \overline{P}_i }{ d_t } \nonumber \\
    & = \frac{ \sum\limits_{\tau = \tau_t}^{\tau_t + \tau^\text{SU}_{t,i} - 1} (\tau - \tau_t) RU_i \Delta \tau }{ d_t } + \frac{ \sum\limits_{\tau = \tau_t + \tau^\text{SU}_{t,i}}^{\tau_{t+1} - 1} \overline{P}_i - \underline{P}_i }{ d_t } + \underline{P}_i \nonumber \\
    & = \frac{{\tau^\text{SU}_{t,i}}}{{{d_t}}}R{U_i}\Delta \tau\frac{{\tau^\text{SU}_{t,i} - 1}}{2} + (1 - \frac{{\tau^\text{SU}_{t,i}}}{{{d_t}}})(\overline{P}_i  - \underline{P}_i ) + \underline{P}_i \nonumber \\
    & = SU_i^{(t)}  \label{eq:trans-rampsu}
\end{align}

Equation \eqref{eq:trans-rampsu} suggests that, if the unit is offline in $t-1$ and becomes online in $t$, $p_i^{(t)}$ reaches its maximum when the power output increases at its ramping up rate $RU_i$ in the original time periods of $t$ from its minimum $\underline{P}_i$ to its maximum $\overline{P}_i$ or to the end of $t$. It is worth mentioning that $\overline{P}_i$ is an upper bound of $SU_i^{(t)}$, so the third parameter of \eqref{eq:conv-con-rampup} remains unchanged.

\section{Case Study}  \label{sec:case}

\subsection{Basic setup}

In this section, the proposed method is verified using the IEEE 118-bus power system and the Polish 2736-bus power system to calculate the day-ahead NCUC. To validate the flexible temporal resolution considering congestion, several methods are utilized here for comparison:

\textbf{M1:} The proposed method, including the flexible temporal resolution determination strategy in Section~\ref{sec:aggre}, the NCUC model in Section~\ref{sec:uc} and a correction strategy in \cite{yu_adaptive_2022}.

\textbf{M2:} A simplified version of the proposed method \cite{yu_adaptive_2022}, where the influence of network constraints in the determination of flexible temporal resolution is neglected.

\textbf{M3:} A method proposed in \cite{pineda_time-adaptive_2019} where a hierarchical clustering strategy is adopted to calculate the starting points of adaptive time periods.

\textbf{M4:} A method where the time periods are evenly merged without considering any feature of the demand or congestion.

\textbf{BM:} The benchmark method using original time periods without simplification.

Despite different strategies for determining flexible temporal resolution, the NCUC model and the correction strategy in the four methods are consistent. The simulations are conducted using Gurobi 9.5.2 \cite{noauthor_gurobi_nodate} on a desktop computer with Intel i9-13900K CPU and 128 GB of RAM.

\subsection{Performance of flexible temporal resolution determination}

\begin{figure}
    \centering
    \includegraphics[width=0.45\textwidth]{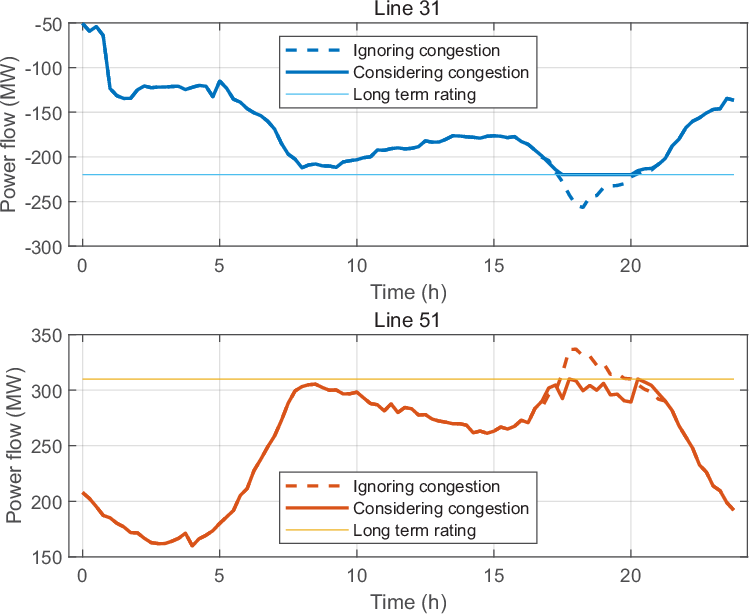}
    \caption{Power flow on Line 31 and Line 51 ignoring and considering congestion. These two lines become congested at approximately 18:00, when the demand is at its highest.}
    \label{fig:case_branch}
\end{figure}

First, the proposed deterministic method of flexible temporal resolution is applied to the 118-bus power system case. The 118-bus power system case includes 118 buses, 54 thermal units and 186 transmission lines, the detailed parameters of which are listed in Table~\ref{tab:param-base}. Based on the UC results ignoring the network constraints, Line 31 and Line 51 are selected to create congestion by decreasing their long-term ratings to 220 MW and 310 MW respectively, as illustrated in Fig.~\ref{fig:case_branch}.

\begin{table}
    \centering
    \caption{Parameters of the 118-bus power system case}    \label{tab:param-base}
    \begin{tabular}{lr}
    \toprule
        Name & Value \\
    \midrule
        Number of buses $N_B$ & 118 \\
        Number of thermal units $N_G$ & 54 \\
        Number of lines $N_L$ & 186 \\
        Number of congested lines & 2 \\
        Total capacity of installed units (MW) & 9966.2 \\
        Minimum on/off time of units (h) & 2.0 \\
        Number of original time periods $T$ & 96 \\
        Number of adaptive time periods $N$ & 38 \\
        Convergence Gap & $10^{-4}$ \\
    \bottomrule
    \end{tabular}
\end{table}

\begin{figure}
    \centering
    \includegraphics[width=0.45\textwidth]{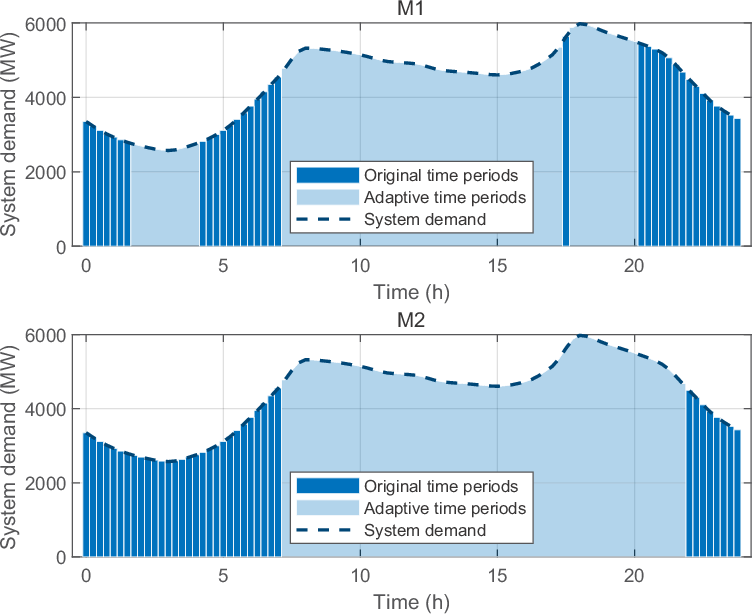}
    \caption{The flexible temporal resolution determined by M1 and M2.}
    \label{fig:case_aggre}
\end{figure}

The results of flexible temporal resolution determination are illustrated in Fig.~\ref{fig:case_aggre}. Here, the light blue areas represent the adaptive time periods composed of multiple original time periods, and the deep blue bars represent the remaining adaptive time periods, each containing one original time period. The results of the proposed method M1 show that in general, the time periods with relatively flat demand curves are likely to be merged and that those with rapid changes in demand tend to remain the same, which essentially agrees with the prerequisite of flexible temporal resolution. In addition, compared with M2, the influence of congestion is reflected at 17:30, where the time periods are split rather than merged into one. The congestion on the two lines sets constraints on the load-following ability of the committed units, which increases the impact brought on the generation side by demand variation and finally affects the temporal resolution.

\subsection{Test results based on the 118-bus power system}

\begin{figure}
    \centering
    \includegraphics[width=0.45\textwidth]{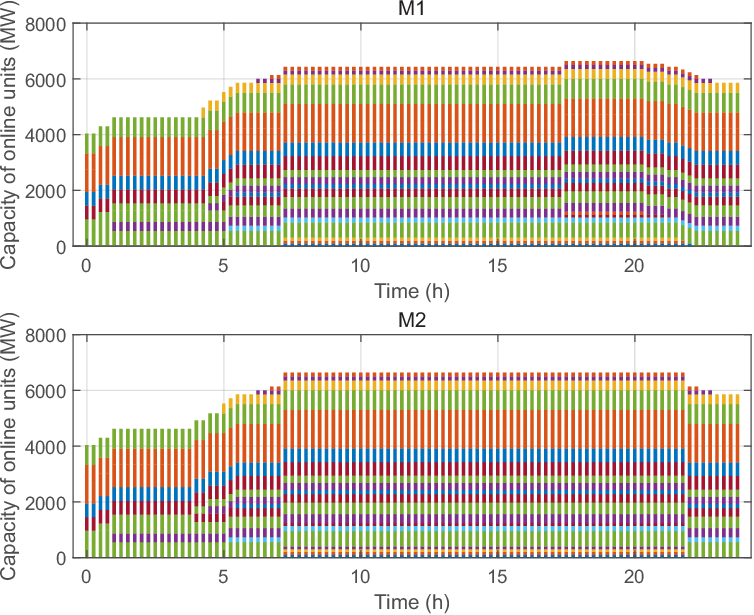}
    \caption{Commitment statuses and online capacities of thermal units of M1 and M2.}
    \label{fig:case_onoff}
\end{figure}

Based on the flexible temporal resolution, NCUC models are optimized to obtain the commitment results for the 118-bus power system. The commitment statuses and capacities of thermal units are represented in Fig.~\ref{fig:case_onoff}. Here, the UC results before correction are visualized. It can be seen that the total online capacity generally reflects the demand level of each time period. By comparing the UC results of M1 and M2, the aforementioned difference in adaptive time periods at 17:30 leads to the difference in commitment statuses. The total capacity of committed units at 17:30 remains the same in M2 but increases in M1 to accommodate for the influence of congestion.

The NCUC results of all the methods are listed in Table~\ref{tab:result-base}. In this case, the NCUC results directly obtained from M2, M3 and M4 are infeasible concerning the original time period sequence, and the second rows of these methods in the table represent the calculation results after correction. All the percentage numbers in parentheses represent relative values with respect to the benchmark, including relative time consumption and relative cost variation. The calculation results show that the proposed method M1 is capable of obtaining a UC solution with low cost variation and few different on/off values. In addition, the time consumption of M1 is merely 1/12.7 of the benchmark, indicating its high computational efficiency. Compared with that of the three other methods, the time used for calculation is similar before correction, as the numbers of integer variables are the same. However, the effectiveness of flexible temporal resolution makes the UC results of M1 free from further correction, thus enhancing its performance.

\begin{table}
    \centering
    \caption{Calculation results of the 118-bus power system case}    \label{tab:result-base}
    \begin{tabular}{lrrr}
    \toprule
        Method & Time (s) & Cost (M\$) & Different on/off statuses \\
    \midrule
        M1 & 9.8(7.9\%) & 3.5047(0.19\%) & 66 \\
        M2 & 11.0(8.8\%) & / & 112 \\
         & 16.8(13.5\%) & 3.5052(0.20\%) & 97 \\
        M3 & 10.5(8.4\%) & / & 17 \\
         & 27.3(21.9\%) & 3.4981(0\%) & 0 \\
        M4 & 9.5(7.6\%) & / & 28 \\
         & 29.8(23.9\%) & 3.4981(0\%) & 0 \\
        BM & 124.7 & 3.4981 & / \\
    \bottomrule
    \end{tabular}
\end{table}

\subsection{Performance with various demand profiles}    \label{subsec:case-demand}

To evaluate the proposed method under various demand scenarios, realistic demand profiles of several provinces in China in a year are combined with the 118-bus power system. A total of 971 days of the annual demand data of several provinces are selected so that the computation time using the original time periods exceeds 60 s for each case, and the maximum system demand in each case equals around 60\% of the total capacity of the installed units.

\begin{figure}
    \centering
    \includegraphics[width=0.45\textwidth]{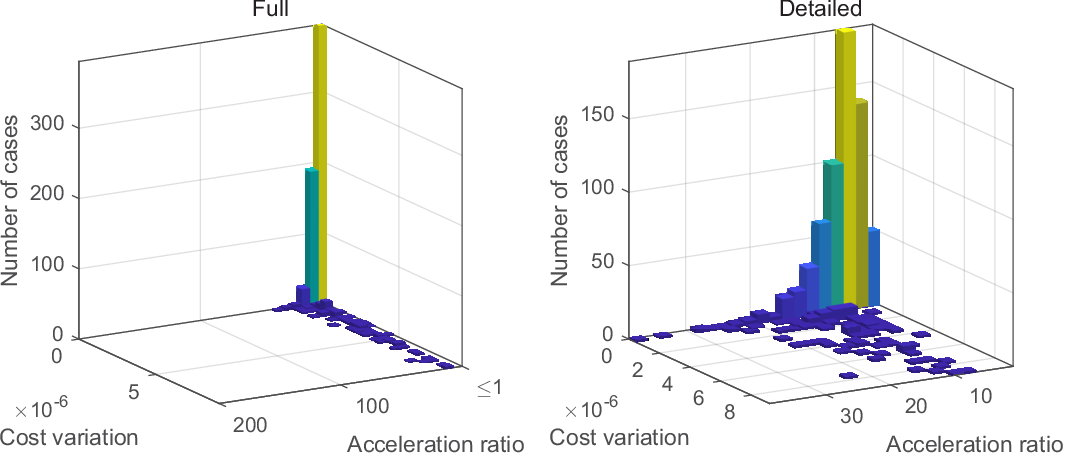}
    \caption{Distribution of acceleration ratios and relative cost variations of UC results with various demand profiles.}
    \label{fig:case_dist}
\end{figure}

The test results are illustrated in Fig.~\ref{fig:case_dist}, which depicts the distribution of the acceleration ratios and the relative cost variations. The majority of acceleration ratios in these cases lie within $3\times$ to $30\times$ (900 cases), with an average acceleration ratio of $8.53\times$. From the perspective of accuracy, the calculation results of most of the cases obtain a relative cost variation below 0.05\%, and only 56 cases exceed 0.1\%.

\begin{figure}
    \centering
    \includegraphics[width=0.45\textwidth]{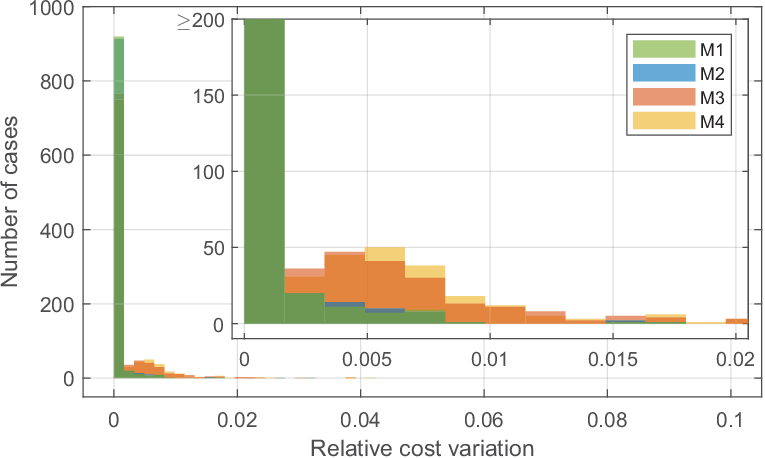}
    \caption{Distribution of relative cost variations of NCUC results with various demand profiles of all the methods.}
    \label{fig:case_dist_comp}
\end{figure}

Applying all the other methods for comparison, the UC results are summarized in Table~\ref{tab:result-multi}, which includes the average values of the three metrics above and the number of cases in need of correction. It can be inferred that the accuracy of the UC results obtained by M1 and M2 is significantly higher than that of the other methods, with lower relative cost variation, fewer different on/off statuses and fewer cases with correction. In addition, the accuracy of the proposed method is better than that of M2 regarding all the metrics due to its consideration of congestion. Besides, the advantage of computational efficiency brought by a reduced number of time periods is basically maintained for M1 as well as the other methods. Fig.~\ref{fig:case_dist_comp} shows a visualized comparison of the cost variation distribution of the cases for different methods, which indicates that there exist more cases of M1 with low cost variation and fewer cases with high cost variation compared with the other three methods, which verifies the economy and the accuracy of the proposed method.

\begin{table}
    \centering
    \caption{Calculation results with various demand profiles}    \label{tab:result-multi}
    \begin{tabular}{lrrrr}
    \toprule
        Method & Acceleration & Cost & Different & Cases with \\
        & ratio & variation (\%) & on/off statuses & correction \\
    \midrule
        M1 & 8.53 & 0.034 & 17.26 & 53 \\
        M2 & 9.90 & 0.037 & 18.16 & 58 \\
        M3 & 12.99 & 0.160 & 63.09 & 298 \\
        M4 & 17.35 & 0.184 & 66.75 & 361 \\
    \bottomrule
    \end{tabular}
\end{table}

\subsection{Comparison of different numbers of time periods}

A significant parameter related to the performance of the proposed method is the number of adaptive time periods $N$, which largely determines the tendency in the balance of efficiency and accuracy. This section provides a comparison of the performance of M1 on the 118-bus system when changing the number of adaptive time periods.

\begin{figure}
    \centering
    \includegraphics[width=0.45\textwidth]{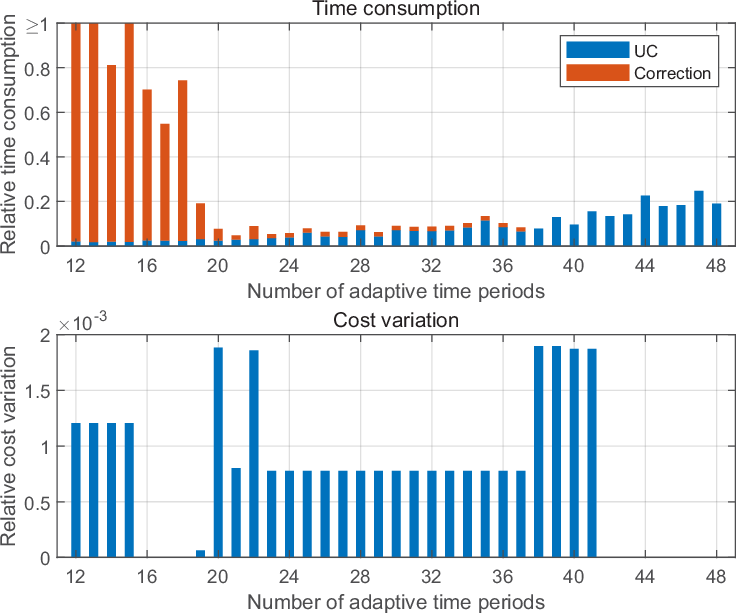}
    \caption{Time consumption and cost variation of M1 with respect to the number of adaptive time periods.}
    \label{fig:case_number}
\end{figure}

Assuming $N$ varies from 12 to 48, the calculation results of M1 on the 118-bus system are depicted in Fig.~\ref{fig:case_number}, including time consumption and cost variation. The time consumption of the NCUC calculation generally decreases with fewer adaptive time periods, which results from the direct shrinkage of the model scale. However, fewer adaptive time periods can also lead to the loss of accuracy of the UC results, as less detail is included and characterized with coarser temporal resolution. When the number of adaptive time periods is too small (in this case, below 38), the NCUC results become infeasible for the original time period sequence, so an extra amount of time becomes necessary for correction, which compromises its efficiency. For the 118-bus power system case, $N=38$ is the smallest number of adaptive time periods without correction.

\subsection{Large-scale test results based on the Polish power system}

To conclude the case study, the Polish 2736-bus power system is utilized for further verification of the proposed method. The Polish power system case includes 2736 buses, 289 thermal units and 3269 transmission lines, whose detailed parameters can be found in Table~\ref{tab:param-polish}. Same as Sec.~\ref{subsec:case-demand}, a total of 17 days of the annual demand data of several provinces in China are selected with each one's computation time using the original time periods exceeding 1000 s.

\begin{table}
    \centering
    \caption{Parameters of the Polish power system case}    \label{tab:param-polish}
    \begin{tabular}{lr}
    \toprule
        Name & Value \\
    \midrule
        Number of buses $N_B$ & 2736 \\
        Number of thermal units $N_G$ & 289 \\
        Number of lines $N_L$ & 3269 \\
        Number of congested lines & 30 \\
        Total capacity of installed units (GW) & 28.88 \\
        Minimum on/off time of units (h) & 6.0 \\
        Number of original time periods $T$ & 96 \\
        Number of adaptive time periods $N$ & 54 \\
        Convergence Gap & $10^{-3}$ \\
    \bottomrule
    \end{tabular}
\end{table}

The NCUC results of all the methods are listed in Table~\ref{tab:result-polish-60}. While the average time consumption of the BM reaches 50 min (120 min maximum), the simplification of M1 leads to an average acceleration ratio of $10.16\times$, and the cost variation of M1 is kept at a low level (0.07\%). In addition, the average time consumption for determining flexible temporal resolution is 1.7 s, which is negligible against the optimization process of the NCUC model. Compared with that of the other methods, the acceleration performance is similar, with M1 being slightly better. However, the results obtained from M1 possess the highest accuracy and economy regarding all the metrics.

To test the performance of the proposed method with higher demand levels, the calculation process is performed again on the large-scale cases above, with the maximum system demand increasing from around 60\% to around 80\%. The NCUC results are listed in Table~\ref{tab:result-polish-80}, where the proposed method achieves an average acceleration ratio of $14.32\times$ and a cost variation of 0.04\%. Compared with the other methods, M1 achieves the best performance in terms of most of the metrics. Therefore, the proposed method is capable of providing a near-optimal NCUC solution while significantly enhancing its computational efficiency, and our method outperforms the other methods in terms of accuracy and economy.

\begin{table}
    \centering
    \caption{Calculation results of the Polish power system case (maximum demand at 60\% of installed capacity)}    \label{tab:result-polish-60}
    \begin{tabular}{lrrrr}
    \toprule
        Method & Acceleration & Cost & Different & Cases with \\
        & ratio & variation (\%) & on/off statuses & correction \\
    \midrule
        M1 & 10.16 & 0.066 & 571.4 & 5 \\
        M2 & 9.10 & 0.097 & 638.1 & 7 \\
        M3 & 9.72 & 0.097 & 648.4 & 12 \\
        M4 & 8.66 & 1.159 & 1388.7 & 16 \\
    \bottomrule
    \end{tabular}
\end{table}

\begin{table}
    \centering
    \caption{Calculation results of the Polish power system case (maximum demand at 80\% of installed capacity)}    \label{tab:result-polish-80}
    \begin{tabular}{lrrrr}
    \toprule
        Method & Acceleration & Cost & Different & Cases with \\
        & ratio & variation (\%) & on/off statuses & correction \\
    \midrule
        M1 & 14.32 & 0.035 & 623.3 & 0 \\
        M2 & 11.65 & 0.032 & 676.0 & 3 \\
        M3 & 8.27 & 0.038 & 627.4 & 3 \\
        M4 & 8.90 & 0.481 & 1049.0 & 10 \\
    \bottomrule
    \end{tabular}
\end{table}

\section{Conclusion}    \label{sec:conclusion}

In this paper, a simplification method for NCUC is proposed to determine the flexible temporal resolution to reduce the computational burden. To maintain the economy and accuracy of the NCUC results as much as possible, the flexible temporal resolution is determined by analyzing a power output variation model of thermal units considering the effect of congestion, which quantifies and utilizes the conception of the impact brought on the thermal units caused by demand variation. The NCUC model is then formulated and modified based on the existing model to adapt to the flexible temporal resolution with adaptive time periods of different durations, and the ramping constraints are transformed by examining the possible extreme circumstances concerning the original time periods. Numerical results validate that the proposed method is capable of achieving an acceleration ratio of $3\times$ to $30\times$ with cost variation mostly under 0.1\% regarding the benchmark with the original time period sequence. Compared with that of the existing methods, by effectively considering the influence of congestion in flexible temporal resolution determination, the overall economy and accuracy of the proposed method are higher, indicating better performance. 

In future work, an efficient correction strategy of infeasible NCUC results obtained by the proposed method will be investigated. We plan to further extend the proposed methodologies in long-term power system planning~\cite{zhong_hierarchical_2022} and electricity market simulation~\cite{chen_security-constrained_2022}, while more similar operational patterns can be collected in these cases to expand the potential computational speedup.

\appendix[Derivation of the Load-following Conditions]

This section presents the derivation of the load-following conditions \eqref{eq:theor-require} based on the power output variation model \eqref{eq:theor-nrange}-\eqref{eq:theor-ndpf}.

\begin{figure}
    \centering
    \includegraphics[width=0.45\textwidth]{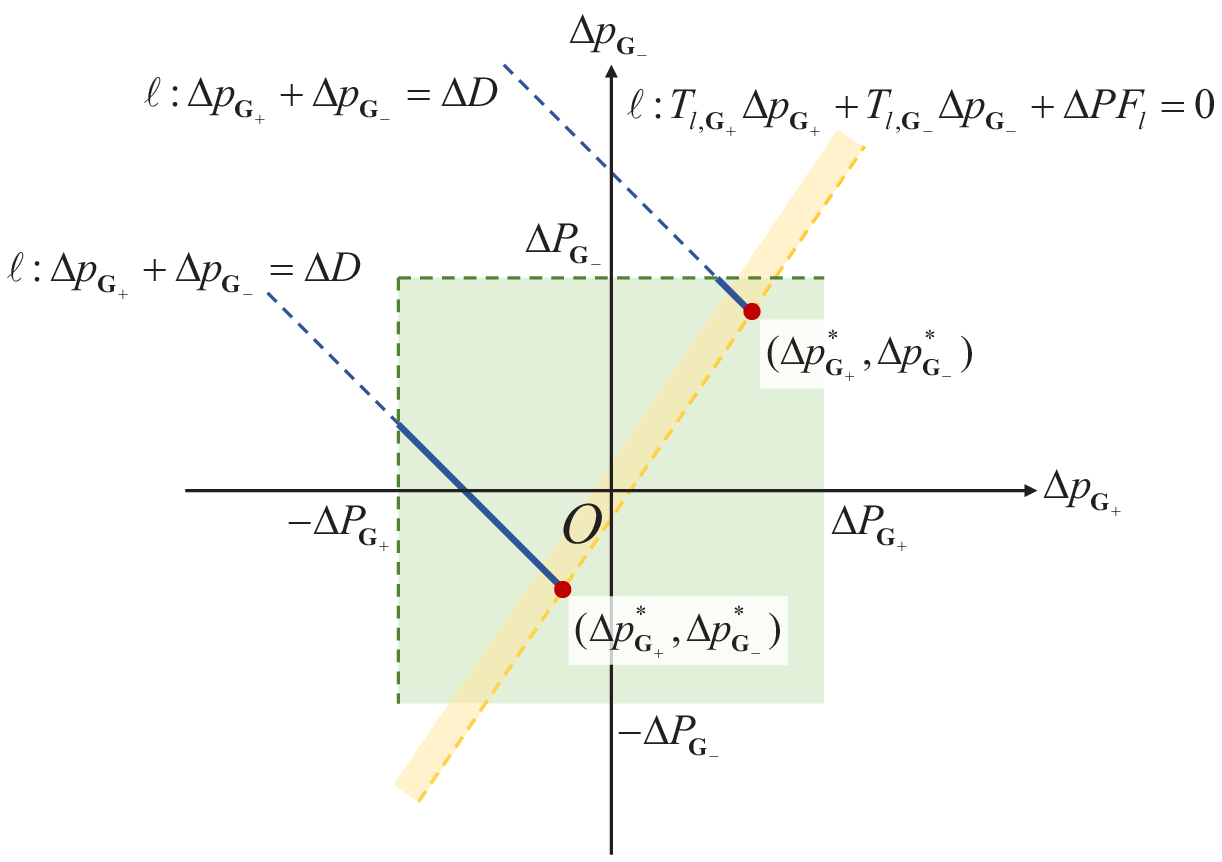}
    \caption{Illustration of the power output variation model on a 2-D plot.}
    \label{fig:append}
\end{figure}

Equations \eqref{eq:theor-nrange}-\eqref{eq:theor-ndpf} consist of only two variables: $\Delta p_{\mathbf{G}_+}$ and $\Delta p_{\mathbf{G}_-}$, and a 2-D $\Delta p_{\mathbf{G}_+}-\Delta p_{\mathbf{G}_-}$ plot is utilized here for visualization. As illustrated in Fig.~\ref{fig:append}, the feasible region of \eqref{eq:theor-nrange} is represented with a square area (green). Equation \eqref{eq:theor-nbal} is represented with a blue line whose slope is equal to -1. For the half-plane corresponding to \eqref{eq:theor-ndpf}, judging from the fact that $T_{l, \mathbf{G}_+} \ge 0$ and $T_{l, \mathbf{G}_-} \le 0$, its edge is represented with a yellow line with a positive slope.

To ensure the feasibility of \eqref{eq:theor-nrange}-\eqref{eq:theor-ndpf}, the aforementioned square area, line and half-plane must intersect with each other. To achieve this goal, the blue line \eqref{eq:theor-nbal} and the half-plane \eqref{eq:theor-ndpf} are first merged into a ray whose terminal point is found by combining their equations:
\begin{equation}
    \begin{cases}
        \Delta p^*_{\mathbf{G}_+} = - \frac{T_{l, \mathbf{G}_-} \Delta D + \Delta PF_l}{T_{l, \mathbf{G}_+} - T_{l, \mathbf{G}_-}} \\
        \Delta p^*_{\mathbf{G}_-} = \frac{T_{l, \mathbf{G}_+} \Delta D + \Delta PF_l}{T_{l, \mathbf{G}_+} - T_{l, \mathbf{G}_-}}
    \end{cases}
\end{equation}

Therefore, ensuring the intersection of the three regions is transformed into ensuring the intersection of the ray and the square area for any $\tau_S \le \tau_s < \tau_f \le \tau_F$, which is equivalent to:
\begin{equation}
    \begin{cases}
        \Delta P_{\mathbf{G}_+} + \Delta P_{\mathbf{G}_-} \ge \Delta D  \\
        - \Delta P_{\mathbf{G}_+} - \Delta P_{\mathbf{G}_-} \le \Delta D    \\
        \Delta p^*_{\mathbf{G}_+} \ge - \Delta P_{\mathbf{G}_+} \\
        \Delta p^*_{\mathbf{G}_-} \le \Delta P_{\mathbf{G}_-}
    \end{cases}, \forall \tau_S \le \tau_s < \tau_f \le \tau_F
\end{equation}

The load-following conditions \eqref{eq:theor-require} can therefore be derived.

\ifCLASSOPTIONcaptionsoff
  \newpage
\fi

\bibliographystyle{IEEEtran}
\bibliography{references.bib}

\begin{IEEEbiography}[{\includegraphics[width=1in,height=1.25in,clip,keepaspectratio]{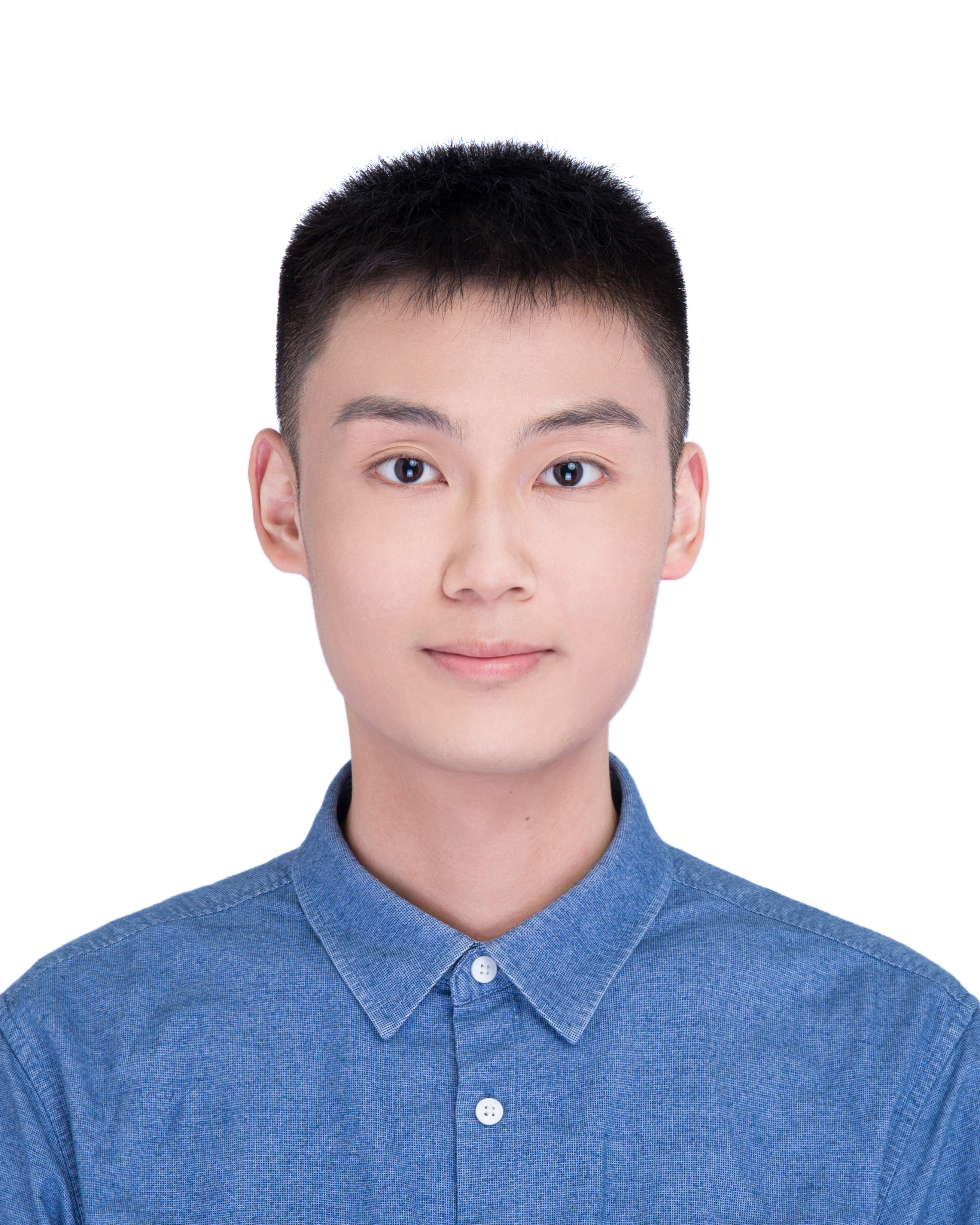}}]{Zekuan Yu}
(Graduate Student Member, IEEE) received the B.S. degree in electrical engineering from Tsinghua University, Beijing, China, in 2022, where he is currently pursuing the Ph.D. degree. His research interests include power system operations and machine learning applications in power systems.
\end{IEEEbiography}

\begin{IEEEbiography}[{\includegraphics[width=1in,height=1.25in,clip,keepaspectratio]{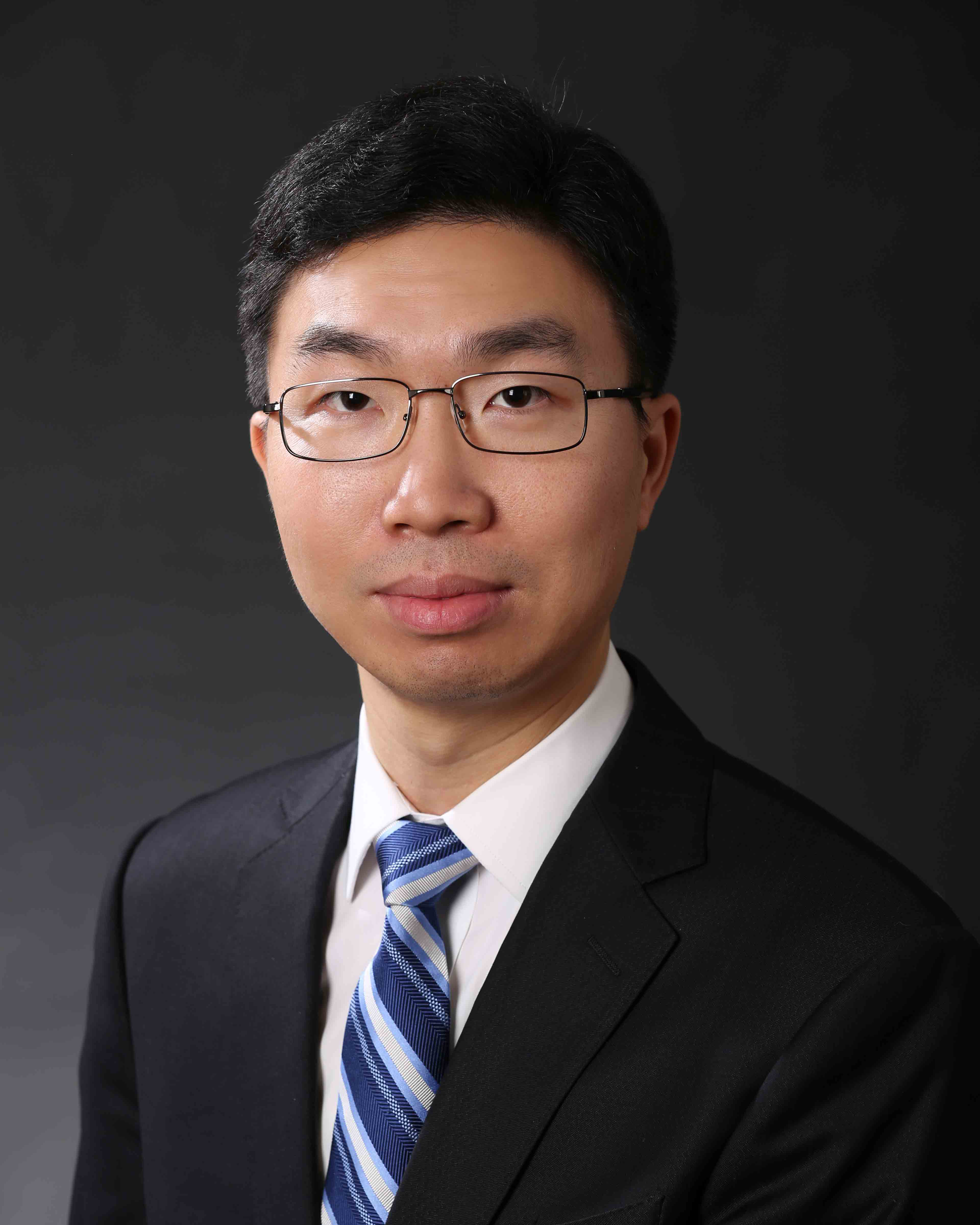}}]{Haiwang Zhong}
(Senior Member, IEEE) received the B.S. and Ph.D. degrees in electrical engineering from Tsinghua University. He is currently an Associate Professor with the Department of Electrical Engineering, Tsinghua University. He is also the Director of Energy Internet Trading and Operation Research Department of Sichuan Energy Internet Research Institute, Tsinghua University. His research interests include power system operations and optimization, electricity markets. He was a recipient of the ProSPER.Net Young Scientist Award and the Outstanding Postdoctoral Fellow of Tsinghua University. He serves as an Associate Editor for the CSEE Journal of Power and Energy Systems. He currently serves as the Secretary of CIGRE D2.53 Working Group on Technology and Applications of Internet of Things in Power Systems.
\end{IEEEbiography}

\begin{IEEEbiography}[{\includegraphics[width=1in,height=1.25in,clip,keepaspectratio]{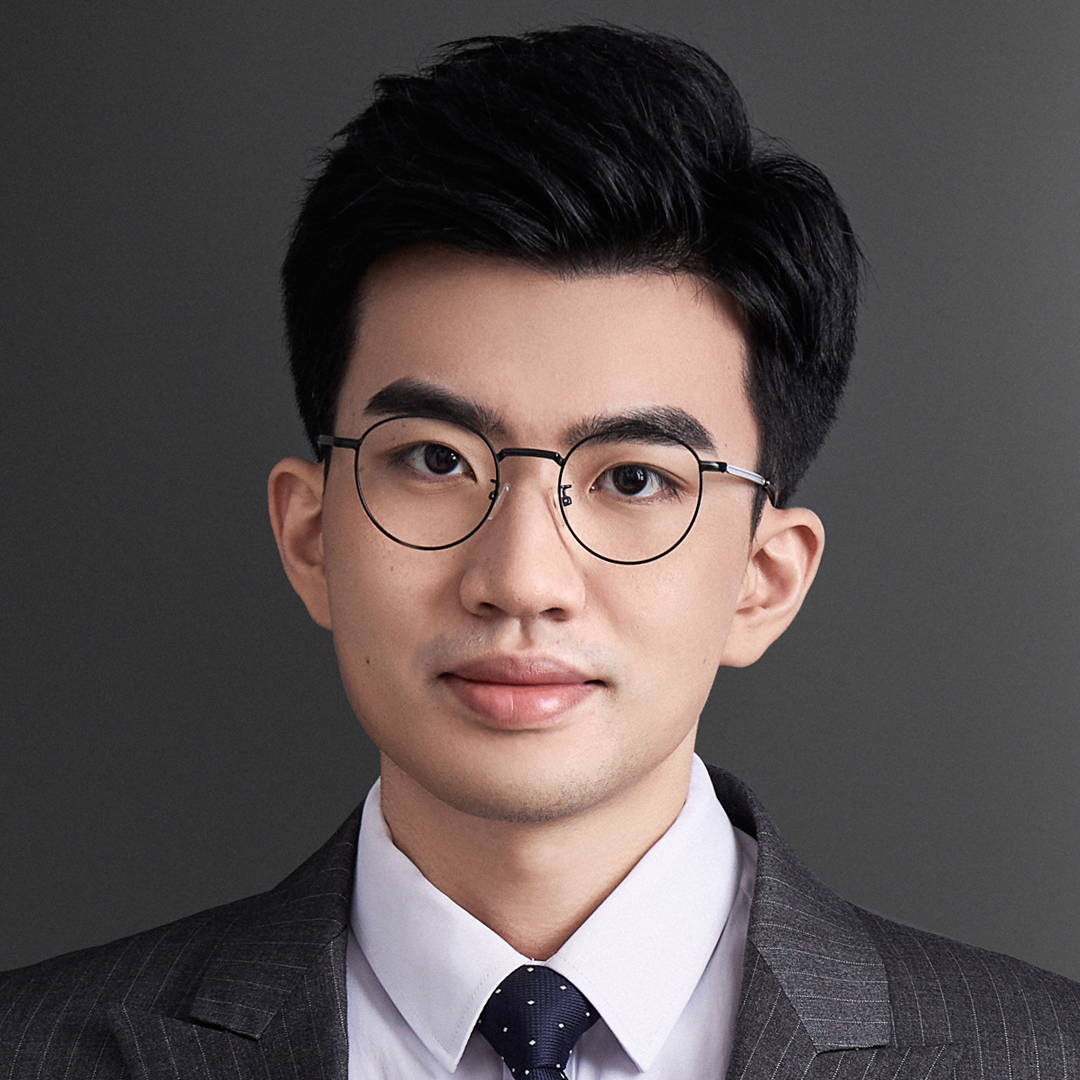}}]{Guangchun (Grant) Ruan}
(Member, IEEE) is currently a postdoc with the Laboratory for Information \& Decision Systems (LIDS) at MIT. Before joining MIT, he received the Ph.D. degree in electrical engineering from Tsinghua University in 2021, and worked as a postdoc with the University of Hong Kong in 2022. He visited Texas A\&M University in 2020 and the University of Washington in 2019. His research interests include electricity market, energy resilience, demand response, data science and machine learning applications.
\end{IEEEbiography}

\begin{IEEEbiography}[{\includegraphics[width=1in,height=1.25in,clip,keepaspectratio]{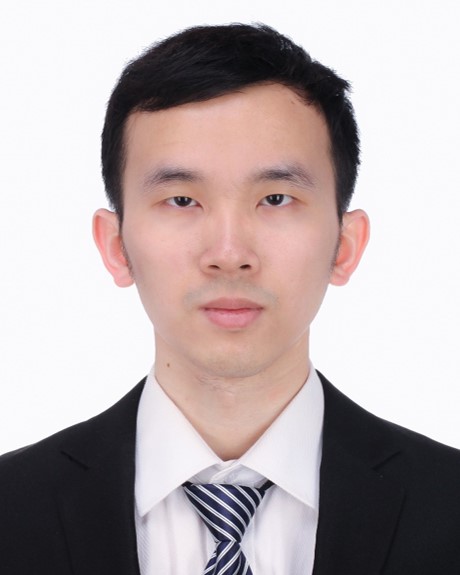}}]{Xinfei Yan}
(Student Member, IEEE) received the B.S. degree in electrical engineering from Tsinghua University, Beijing, China, in 2019, where he is currently pursuing the Ph.D. degree. His research interests include distributed optimization algorithms and power system operations.
\end{IEEEbiography}

\end{document}